\documentstyle[preprint,aps]{revtex}
\tightenlines
\begin{document}
\draft
\title{To the question of a nonrelativistic wave equation \\
for a system of interacting particles}
\author{M. V. Kuzmenko}
\address{Bogolyubov Institute for Theoretical Physics of the NAS of Ukraine, \\
Metrolohichna Str., 14b, Kyiv-143, 252143 Ukraine}
\date{\today}
\maketitle

\begin{abstract}
It is shown that the Schr\"{o}dinger nonrelativistic equation of a system of
interacting particles is not a rigorously nonrelativistic equation since it
is based on the implicit assumption of finiteness of the interaction
propagation velocity. For a system of interacting particles, a fully
nonrelativistic nonlinear system of integro-differential equations is
proposed. In the case where the size of the system of particles is of the
same order as the Compton wavelength associated with particles, certain
essential differences are shown to exist as compared with traditional
consequences of the nonrelativistic Schr\"{o}dinger equation.
\end{abstract}


\section{Introduction}

In classical mechanics, the problem of two particles, which interact with
force depending only on the relative distance between them, is separated
into two three-dimensional problems: the free-particle problem and one of a
particle in a static potential field~\cite{R1}. Motion of a free particle,
whose mass is equal to the sum of the masses of both particles, is a motion
of the center of masses of the system and is uniform and rectilinear. Motion
of a relative particle with so-called reduced mass occurs in the field with
potential $V({\bf r})$.

An analogous situation is also observed for the nonrelativistic wave
equation of a system of two interacting particles, which was proposed by
Schr\"{o}dinger already in the second communication on wave mechanics by the
example of elastic rotator (two-atom molecule)~\cite{R2}.

The probabilistic interpretation of the square of the modulus of a wave
function is possible only under the assumption that measurements of
coordinates or momenta of various particles do not principally disturb one
another even if there exists some interaction between particles~\cite{R3}.
This means that operators of coordinates and momenta of two particles
commute with each other. But in the theory of Schr\"{o}dinger, still
operators of coordinates and momenta of various particles commute with one
another that is equivalent to the lacking of any interference on measurement
of the coordinate of one particle and the momentum of the other. The last
assertion is valid if the duration of measurement of the coordinate of a
particle is much less than the duration of propagation of a light signal at
distances of about the size of the system, or, what is the same, if the
Compton wave length is much less than the size of the system. Therefore, the
Schr\"{o}dinger equation perfectly works in atomic physics and solid-state
physics. But a direct application of the Schr\"{o}dinger equation to atomic
nuclei seems not entirely correct because the Compton wave length of a
nucleon is of order of the size of an atomic nucleus itself. In addition, a
rigorous nonrelativistic statement requires to consider the interaction
propagation velocity to be infinitely large that forces us to assume that
operators of coordinates and momenta of various particles do not commute
with one another.

\section{ A completely nonrelativistic statement of the quantum two-body
problem}

As is known, classical equations of motion of a particle with mass $m$ in an
external field $V({\bf r})$ follow from the Hamilton function
\begin{equation}
H({\bf r,p})=\frac{{\bf p}^{2}}{2m}+V({\bf r})\,,  \label{E1}
\end{equation}
which depends on the coordinates ${\bf r}$ of the particle and on the
corresponding momentum ${\bf p}$. The total energy of the system
\begin{equation}
E=H({\bf r,p})\,.  \label{E2}
\end{equation}

With this classical system, we associate a quantum system, whose dynamic
state is represented by the wave function $\Psi ({\bf r,}t)$ defined in the
configuration space. The wave equation is deduced by the formal substitution
of the quantities $E$, ${\bf r}$, ${\bf p}$ in both sides of relation~(\ref
{E2}) by the corresponding operators~\cite{R4}:
\begin{eqnarray}
&&E\rightarrow \hat{E}=i\hbar \frac{\partial }{\partial t}\,,  \label{E3} \\
&&{\bf r\rightarrow \hat{r}=r\,,}  \label{E4} \\
&&{\bf p\rightarrow \hat{p}=-}i\hbar {\bf \nabla }_{{\bf r}}\,.  \label{E5}
\end{eqnarray}
Here, $\hbar ={\displaystyle{\frac{h}{2\pi }}}$, where $h$ is the universal
constant introduced by Planck.

It is meant that the result of action of both sides of equality~(\ref{E2}),
considered as operators, on $\Psi ({\bf r,}t)$ is the same. The realization
of this fact implies the Schr\"{o}dinger nonrelativistic equation for a
particle in an external field $V({\bf r})$:
\begin{equation}
i\hbar \frac{\partial }{\partial t}\Psi ({\bf r,}t)=\left[ -\frac{\hbar ^{2}%
}{2m}\Delta +V({\bf r})\right] \Psi ({\bf r,}t)\,.  \label{E6}
\end{equation}

It is worth to emphasize that the operators ${\bf \hat{r}}$ and ${\bf \hat{p}%
}$ in~(\ref{E4}), (\ref{E5}) are written in the configuration space, and $%
{\bf r}$ is the vector of position of the particle in the Cartesian
coordinate system.

Operators of coordinate and momentum do not commute with each other:
\begin{equation}
\left[ \hat{x},\,\hat{p}_{x}\right] =i\hbar \,,\text{\quad }\left[ \hat{y},\,%
\hat{p}_{y}\right] =i\hbar \,,\text{\quad }\left[ \hat{z},\,\hat{p}_{z}%
\right] =i\hbar \,,  \label{E7}
\end{equation}
that leads to the Heisenberg uncertainty relations
\begin{equation}
\Delta x\,\Delta p_{x}\geq \hbar /2\,,\text{\quad }\Delta y\,\Delta
p_{y}\geq \hbar /2\,,\text{\quad }\Delta z\,\Delta p_{z}\geq \hbar /2\,,
\label{E8}
\end{equation}
where the quantities $\Delta x$, $\Delta p_{x}$, $\Delta y$, $\Delta p_{y}$,
$\Delta z$, and $\Delta p_{z}$ are directly connected with corresponding
measurements and present mean square deviations from the mean value. For
example, we have
\begin{equation}
\Delta x=\sqrt{\left\langle \hat{x}^{2}\right\rangle -\left\langle \hat{x}%
\right\rangle ^{2}}\,  \label{E9}
\end{equation}
for the coordinate $x$, by definition, where $\left\langle \hat{A}%
\right\rangle $ is the mean value of the operator $\hat{A}$ in the dynamic
state defined by a wave function $\Psi ({\bf r,}t)$.

Relations~(\ref{E8}) assert that a particle cannot be in states, in which
its coordinate and momentum simultaneously take quite definite, exact
values. Moreover, quantum theory accepts that an unpredictable and
uncontrolled perturbation, undergone by a physical system in the process of
measurement, is always finite and such that the Heisenberg uncertainty
relations~(\ref{E8}) are satisfied~\cite{R4}. Hence, no experiment can lead
to a simultaneous exact measurement of the coordinate and momentum of a
particle. For example, the measurement of the coordinate $x$ with accuracy $%
\Delta x$ in the well-known experiment with a microscope, considered by
Heisenberg, is accompanied by the uncontrolled transfer of momentum to the
particle, which is characterized by the uncertainty
\begin{equation}
\Delta p_{x}\sim \frac{\hbar }{2\Delta x}\,.  \label{E10}
\end{equation}
In this case, limits of the accuracy of determination of a position are set
always by the optical resolution stipulated by diffraction effects according
to classical wave optics. For example, it is known that the limit accuracy
of an image $\Delta x$ for a microscope is defined by the formula
\begin{equation}
\Delta x\sim \frac{\lambda ^{\prime }}{\sin \vartheta }\,,  \label{E11}
\end{equation}
where $\lambda ^{\prime }$ is the wave length of scattered light, which can
differ from that of incident light, and $\vartheta $ is half the objective
aperture. According to relation~(\ref{E11}), to increase the accuracy, it is
profitable to have the wave length of scattered light as short as possible.
But owing to the Compton effect, the frequency of scattered light changes by
a value defined by the conservation laws of energy and momentum. This
implies that, even in the limit $\nu \rightarrow \infty $\quad $(\lambda
=c/\nu \rightarrow 0)$, the frequency of scattered emission $\nu ^{\prime }$
cannot exceed some finite value. If ${\bf p}$ is a momentum, ${\bf v}$ is a
velocity, and $E=c\sqrt{m^{2}c^{2}+p^{2}}$ is the energy of a material
particle prior to the process of scattering, then we have~\cite{R3}
\begin{eqnarray}
&&\nu ^{\prime }=\frac{mc^{2}}{h}\frac{1}{\sqrt{1-v^{2}/c^{2}}}\,,
\label{E12} \\
&&\lambda ^{\prime }=\frac{h}{mc}\sqrt{1-v^{2}/c^{2}}\,  \label{E13}
\end{eqnarray}
in this limit, which gives a maximum value for $\nu ^{\prime }$ and a
minimum one for $\lambda ^{\prime }=c/\nu ^{\prime }$. Here, $c$ is the
velocity of light in vacuum.

Thus, to attain the maximum accuracy of determination of a position on the
observation of the scattering of a quantum of light by using an optical
instrument, we obtain the following expression:
\begin{equation}
\Delta x=\frac{h}{mc}\sqrt{1-v^{2}/c^{2}}\,.  \label{E14}
\end{equation}

The duration of the process of measurement of a position, i.e., the time,
during which the interaction between a light quantum and a particle can
occur, can in no way be less than the periods of oscillations of the
incident and scattered emissions and should be of the order of ${1/}\nu
^{\prime }:$
\begin{equation}
\Delta t=\frac{h}{mc^{2}}\sqrt{1-v^{2}/c^{2}}\,.  \label{E15}
\end{equation}

If the size of the system is such that a characteristic time of flight for
the system significantly exceeds $\Delta t$, then one can say that the
process of measurement of the particle coordinate with accuracy $\Delta x$
is followed by an impact on a particle with the force
\begin{equation}
F_{x}\sim \frac{\Delta p_{x}}{\Delta t}\sim \frac{\hbar c}{2(\Delta x)^{2}}%
\,.  \label{E16}
\end{equation}
Here, we assume that the momentum transferred to a particle under
measurement of its coordinate is of order of the mean square deviation $%
\Delta p_{x}$.

On the measurement of the momentum of a particle with accuracy $\Delta p_{x}$%
, it undergoes an impact with force
\begin{equation}
F_{x}\sim \frac{2c}{\hbar }(\Delta p_{x})^{2}\,.  \label{E17}
\end{equation}

Consider now a system of two interacting particles, whose Hamilton function
is
\begin{equation}
H=\frac{{\bf p}_{1}^{2}}{2m_{1}}+\frac{{\bf p}_{2}^{2}}{2m_{2}}+V(\left|
{\bf r}_{2}-{\bf r}_{1}\right| )\,,  \label{E18}
\end{equation}
where ${\bf r}_{1}$ and ${\bf r}_{2}$ are Cartesian coordinates of two
particles with masses $m_{1}$ and $m_{2}$, ${\bf p}_{1}$ and ${\bf p}_{2}$
are their corresponding momenta, and the potential energy depends only on
the distance between particles. To derive the nonrelativistic
Schr\"{o}dinger equation for this system, we proceed analogously to Eq.~(\ref
{E6}).

This classical system is put into correspondence to the quantum system,
whose dynamic state is represented by the wave function $\Psi ({\bf r}_{1}%
{\bf ,r}_{2}{\bf ,}t)$ defined in the configuration space. The wave equation
can be derived by the formal replacement of the quantities $E$, ${\bf r}_{1}$%
, ${\bf r}_{2}$, ${\bf p}_{1}$, and ${\bf p}_{2}$ on both sides of a
relation analogous to~(\ref{E2}) by the relevant operators
\begin{eqnarray}
&&E\rightarrow \hat{E}=i\hbar \frac{\partial }{\partial t}\,,  \label{E19} \\
&&{\bf r}_{1}{\bf \rightarrow \hat{r}}_{1}{\bf =r}_{1}\,,  \label{E20} \\
&&{\bf r}_{2}{\bf \rightarrow \hat{r}}_{2}{\bf =r}_{2}\,,  \label{E21} \\
&&{\bf p}_{1}{\bf \rightarrow \hat{p}}_{1}{\bf =-}i\hbar {\bf \nabla }_{1}\,,
\label{E22} \\
&&{\bf p}_{2}{\bf \rightarrow \hat{p}}_{2}{\bf =-}i\hbar {\bf \nabla }_{2}\,.
\label{E23}
\end{eqnarray}

Then the well-known Schr\"{o}dinger nonrelativistic equation for a system of
two interacting particles has the form
\begin{equation}
i\hbar \frac{\partial }{\partial t}\Psi ({\bf r}_{1}{\bf ,r}_{2}{\bf ,}t)=%
\left[ -\frac{\hbar ^{2}}{2m_{1}}\Delta _{1}-\frac{\hbar ^{2}}{2m_{2}}\Delta
_{2}+V(\left| {\bf r}_{2}-{\bf r}_{1}\right| )\right] \Psi ({\bf r}_{1}{\bf %
,r}_{2}{\bf ,}t)\,.  \label{E24}
\end{equation}

The operators ${\bf \hat{r}}_{1}$, ${\bf \hat{r}}_{2}$, ${\bf \hat{p}}_{1}$,
and ${\bf \hat{p}}_{2}$ are such that they satisfy the following commutation
relations:
\begin{equation}
\left[ \hat{x}_{k},\,\hat{p}_{kx}\right] =i\hbar \,,\text{\quad }\left[
\hat{y}_{k},\,\hat{p}_{ky}\right] =i\hbar \,,\text{\quad }\left[ \hat{z}%
_{k},\,\hat{p}_{kz}\right] =i\hbar \,,\text{\quad }k=1,\,2\,.  \label{E25}
\end{equation}

All other possible commutation relations equal zero, including
\begin{equation}
\left[ \hat{x}_{k},\,\hat{p}_{lx}\right] =0\,,\text{\quad }\left[ \hat{y}%
_{k},\,\hat{p}_{ly}\right] =0\,,\text{\quad }\left[ \hat{z}_{k},\,\hat{p}%
_{lz}\right] =0\,,\text{\quad }k,\,l=1,\,2\,,\text{\quad }k\neq l\,.
\label{E26}
\end{equation}

Equalities~(\ref{E26}) are based on the assumption that measurements of
coordinates and momenta of different particles do not disturb one another in
principle even in the presence of some interaction forces between particles~
\cite{R3}. That is, one supposes that the change in the force action of a
particle on another one, caused by a measurement of the coordinate of the
first, propagates with finite velocity.

Thus, to derive the Schr\"{o}dinger nonrelativistic equation for a
two-particle system, one uses, on the one hand, the Hamilton classical
nonrelativistic function and, on the other hand, the implicit assumption
about finiteness of the interaction propagation velocity.

In the fully nonrelativistic quantum theory, we must consider the
interaction propagation velocity as infinitely large, which forces us to
drop the requirement for the commutation relations~(\ref{E26}) to hold.
Having accepted this viewpoint, we will consider that, under a measurement
of the coordinate of the first particle, there occurs the uncontrolled
transfer of momentum not only to this particle but to the whole system since
the particles are connected through the interaction potential, whose
propagation velocity is infinitely large. Therefore, it is natural to demand
that the commutator of the operator of the coordinate of the first particle
with the operator of the total momentum of the system be equal to $i\hbar $:
\begin{equation}
\left[ \hat{x}_{1},\,\hat{P}_{cx}\right] =i\hbar \,,\text{\quad }%
\left[ \hat{y}_{1},\,\hat{P}_{cy}\right] =i\hbar \,,\text{\quad }%
\left[ \hat{z}_{1},\,\hat{P}_{cz}\right] =i\hbar \,\,. \label{E27}
\end{equation}
Here, ${\bf \hat{P}}_{c}={\bf \hat{p}}_{1}+{\bf \hat{p}}_{2}$ is
the operator of the total momentum of the system. The same should
be true for the second particle:
\begin{equation}
\left[ \hat{x}_{2},\,\hat{P}_{cx}\right] =i\hbar \,,\text{\quad }%
\left[ \hat{y}_{2},\,\hat{P}_{cy}\right] =i\hbar \,,\text{\quad }%
\left[ \hat{z}_{2},\,\hat{P}_{cz}\right] =i\hbar \,. \label{E28}
\end{equation}

Note that relations~(\ref{E27}) and (\ref{E28}) hold true also for a
Schr\"{o}dinger nonrelativistic equation. Namely, they allow one to
construct the operator of coordinates of the center of masses of the system.
The commutator of the last with the operator of the total momentum of the
system equals $i\hbar $. On the contrary, the fulfillment of relations~(\ref
{E25}) is not obligatory for a system of interacting particles, and we
intend to reject this requirement.

It is clear that, on measuring the coordinate of some particle with accuracy
$\Delta x$, the system undergoes an impact with force $\sim \hbar c/2(\Delta
x)^{2}$. For example, the measurement of the coordinate of a nonrelativistic
electron under observation of the scattering of a light quantum with an
optical device with the greatest possible accuracy of order of the Compton
wave length $\lambda _{e}=h/m_{e}c=2.4\cdot 10^{-10}\mathop{\rm cm}$ is
accompanied by impact with the force $F_{e}\sim 10^{8}\mathop{\rm MeV}/%
\mathop{\rm cm}$. For a proton with its Compton wave length of the order of $%
1.3\cdot 10^{-13}\mathop{\rm
cm}$, the impact force is about $F_{p}\sim 10^{15}\mathop{\rm MeV}/%
\mathop{\rm cm}$. The mean interaction force between particles in a hydrogen
atom in the ground state is $F_{\text{H}}\sim 10^{4}\mathop{\rm
MeV}/\mathop{\rm cm} $, and that for the bound state of the nucleus of
deuterium is $F_{\text{D}}\sim 10^{14}\mathop{\rm MeV}/\mathop{\rm cm}$.
Therefore, whereas we can neglect the interaction force $F_{\text{H}%
}/F_{e}\sim 10^{-4}$ between particles on measuring the coordinates of
particles in an atom and consider the operators of coordinates and momenta
of various particles to be commuting, it is not the case for an atomic
nucleus, because the ratio of the interparticle interaction force to the
impact one is of order $F_{\text{D}}/F_{p}\sim 10^{-1}$.

In the general case, let
\begin{equation}
\left[ \hat{x}_{1},\,\hat{p}_{2x}\right] =i\hbar \hat{f}_{1}\,,  \label{E29}
\end{equation}
where $\hat{f}_{1}$ is some dimensionless Hermitian operator. Then it
follows from Eq.~(\ref{E27}) that
\begin{equation}
\left[ \hat{x}_{1},\,\hat{p}_{1x}\right] =i\hbar (1-\hat{f}_{1})\,.
\label{E30}
\end{equation}

By analogy, if
\begin{equation}
\left[ \hat{x}_{2},\,\hat{p}_{1x}\right] =i\hbar \hat{f}_{2}\,,  \label{E31}
\end{equation}
then
\begin{equation}
\left[ \hat{x}_{2},\,\hat{p}_{2x}\right] =i\hbar (1-\hat{f}_{2})\,.
\label{E32}
\end{equation}

The dimensionless Hermitian operators $\hat{f}_{1}$ and $\hat{f}_{2}$ depend
generally on the interaction force between particles ${\bf F}_{12}$ and on
masses of the interacting particles $m_{1}$ and $m_{2}$. The operators $\hat{%
f}_{1}$ and $\hat{f}_{2}$ cannot depend on a direction of the vector ${\bf F}%
_{12}$, since the commutation relations for the $x$, $y$, and $z$ components
should be identical by analogy with~(\ref{E29})-(\ref{E32}), since there are
no separated directions in the system, and independent variables in the
Cartesian coordinate system are fully equivalent. For this reason, the
operators $\hat{f}_{1}$ and $\hat{f}_{2}$ are only functions of the absolute
value of a force, i.e., of $F_{12}^{2}$:
\begin{equation}
\hat{f}_{1}\equiv \hat{f}_{1}(m_{1},m_{2},F_{12}^{2})\,,\text{\quad }\hat{f}%
_{2}\equiv \hat{f}_{2}(m_{1},m_{2},F_{12}^{2})\,.  \label{E33}
\end{equation}

Let us make permutation of $m_{1}$ and $m_{2}$. Then
\begin{equation}
\left[ \hat{x}_{1},\,\hat{p}_{2x}\right] =i\hbar \hat{f}%
_{1}(m_{2},m_{1},F_{12}^{2})\,,\text{\quad }\left[ \hat{x}_{2},\,\hat{p}_{1x}%
\right] =i\hbar \hat{f}_{2}(m_{2},m_{1},F_{12}^{2})\,.  \label{E34}
\end{equation}
Compare~(\ref{E34}) with~(\ref{E29}), (\ref{E31}). Considering that the
physical situation has not changed, we get
\begin{equation}
\hat{f}_{1}(m_{2},m_{1},F_{12}^{2})=\hat{f}_{2}(m_{1},m_{2},F_{12}^{2})\,.
\label{E35}
\end{equation}

Thus, we have one unknown operator $\hat{f}_{1}(m_{1},m_{2},F_{12}^{2})$.
For $m_{2}\rightarrow 0$, $\hat{f}_{1}$ must tend to zero since, in the
absence of the second particle, the whole momentum transferred under the
measurement of the coordinate $x_{1}$ falls namely this particle. If $%
F_{12}\rightarrow 0$, then $\hat{f}_{1}\rightarrow 0$, i.e., without any
interaction forces between particles, the operators of coordinates and
momenta of different particles commute among themselves. The situation $%
F_{12}\rightarrow \infty $ corresponds to the case where we have one
particle of mass $M$ and mentally represent that it consists of two strongly
bound particles with masses $m_{1}$ and $m_{2}$. Therefore, the momentum,
received under a measurement of some coordinate, is distributed
proportionally to masses of particles. This enables us to write down $\hat{f}%
_{1}$ as $\hat{f}_{1}=m_{2}/M$. Here, $M=m_{1}+m_{2}$ is the system mass.

Therefore, without loss of generality, we can present the operator $\hat{f}%
_{1}$ as
\begin{equation}
\hat{f}_{1}=\frac{m_{2}}{M}\hat{\varepsilon}(F_{12}^{2},m_{1},m_{2})\,,
\label{E36}
\end{equation}
where $\hat{\varepsilon}$ is a new operator, which is assumed to be
symmetric with respect to the masses of particles $m_{1}$ and $m_{2}$. In
what follows, we will omit its explicit dependence on masses to shorten
formulas, namely, $\hat{\varepsilon}(F_{12}^{2},m_{1},m_{2})\equiv \hat{%
\varepsilon}(F_{12}^{2})$. For $F_{12}\rightarrow 0$, $\hat{\varepsilon}%
\rightarrow 0$, and $\hat{\varepsilon}\rightarrow 1$ for $F_{12}\rightarrow
\infty $.

For the noncommuting operators $\hat{x}_{1}$ and $\hat{p}_{2x}$, the
uncertainty relation has the form~\cite{R5}
\begin{equation}
\Delta x_{1}\,\Delta p_{2x}\geq \frac{\hbar }{2}\frac{m_{2}}{M}\left|
\left\langle \hat{\varepsilon}(F_{12}^{2})\right\rangle \right| \,,
\label{E37}
\end{equation}
where $\left\langle \hat{\varepsilon}(F_{12}^{2})\right\rangle \equiv
\varepsilon $ is a quantum-mechanical average in the state $\Psi ({\bf r}_{1}%
{\bf ,r}_{2}{\bf ,}t)$.

If we replace the operator $\hat{\varepsilon}$ in~(\ref{E36}) by its
averaged quantum-mechanical value $\left\langle \hat{\varepsilon}%
(F_{12}^{2})\right\rangle $, the uncertainty relation~(\ref{E37}) does not
change. This makes it possible to write down a nonrelativistic wave equation
for a two-particle system, since the operator $\hat{f}_{1}$ is now a
constant.

It is worth to say several words about the commutativity of the operators of
coordinates and momenta of different particles between themselves:
\begin{equation}
\left[ \hat{x}_{1},\,\hat{x}_{2}\right] =0\,,\text{\quad }\left[ \hat{p}%
_{1x},\,\hat{p}_{2x}\right] =0\,.  \label{E38}
\end{equation}
If we increase the accuracy of measurements of the coordinates $x_{1}$, $%
x_{2}$, the impact forces $F_{1}=\hbar c/2(\Delta x_{1})^{2}$ and $%
F_{2}=\hbar c/2(\Delta x_{2})^{2}$ also grow. For $\Delta x_{1}\rightarrow 0$
and $\Delta x_{2}\rightarrow 0$, we have $F_{1}\gg F_{12}$, $F_{2}\gg F_{12}$%
. Therefore, we can neglect the interaction force $F_{12}$ between particles
and consider the operators of coordinates of both particles as commuting.
Analogously, on measuring the momenta $p_{1x}$, $p_{2x}$, the impact forces $%
F_{1}=2c(\Delta p_{1x})^{2}$ $/\hbar $ and $F_{2}=2c(\Delta
p_{2x})^{2}/\hbar $ tend to zero on increasing the accuracy of measurements.
For this reason, the operators $\hat{p}_{1x}$ and $\hat{p}_{2x}$ also can be
considered as commuting.

We present now the commutation relations for all operators of coordinates
and momenta in the two-body problem:
\begin{eqnarray}
&&\left[ \hat{x}_{1},\,\hat{p}_{1x}\right] =i\hbar \left( 1-\frac{m_{2}}{M}%
\varepsilon \right) \,,  \label{E39} \\
&&\left[ \hat{x}_{2},\,\hat{p}_{2x}\right] =i\hbar \left( 1-\frac{m_{1}}{M}%
\varepsilon \right) \,,  \label{E40} \\
&&\left[ \hat{x}_{1},\,\hat{p}_{2x}\right] =i\hbar \frac{m_{2}}{M}%
\varepsilon \,,  \label{E41} \\
&&\left[ \hat{x}_{2},\,\hat{p}_{1x}\right] =i\hbar \frac{m_{1}}{M}%
\varepsilon \,,  \label{E42} \\
&&\left[ \hat{x}_{1},\,\hat{x}_{2}\right] =0\,,  \label{E43} \\
&&\left[ \hat{p}_{1x},\,\hat{p}_{2x}\right] =0\,.  \label{E44}
\end{eqnarray}

For the $y$ and $z$ components, we have analogous relations. We recall that $%
\varepsilon $ is a quantum mechanical mean value of the operator $\hat{%
\varepsilon}(F_{12}^{2})$ in the state $\Psi ({\bf r}_{1}{\bf ,r}_{2}{\bf ,}%
t)$:
\begin{equation}
\varepsilon =\frac{\left\langle \Psi ,\hat{\varepsilon}(F_{12}^{2})\Psi
\right\rangle }{\left\langle \Psi ,\Psi \right\rangle }\,.  \label{E45}
\end{equation}

We can construct now one of the possible representations for the operators
of coordinates and momenta of a two-particle system:
\begin{eqnarray}
&&{\bf \hat{r}}_{1}=-(1-\varepsilon )\frac{m_{2}}{M}{\bf r+R\,\,,}
\label{E46} \\
&&{\bf \hat{r}}_{2}=(1-\varepsilon )\frac{m_{1}}{M}{\bf r+R\,\,,}
\label{E47} \\
&&{\bf \hat{p}}_{1}=i\hbar {\bf \nabla }_{{\bf r}}-i\hbar \frac{m_{1}}{M}%
{\bf \nabla }_{{\bf R}}\,,  \label{E48} \\
&&{\bf \hat{p}}_{2}=-i\hbar {\bf \nabla }_{{\bf r}}-i\hbar \frac{m_{2}}{M}%
{\bf \nabla }_{{\bf R}}\,.  \label{E49}
\end{eqnarray}
It is easily to verify that operators~(\ref{E46})-(\ref{E49}) satisfy the
commutation relations~(\ref{E39})-(\ref{E44}).

In (\ref{E46})-(\ref{E49}), ${\bf r}$ and ${\bf R}$ are independent operator
variables. The latter represents coordinates of the center of masses of the
system:
\begin{equation}
{\bf \hat{r}}_{c}=\frac{m_{1}{\bf \hat{r}}_{1}+m_{2}{\bf \hat{r}}_{2}%
}{m_{1}+m_{2}}={\bf R\,\,.}  \label{E50}
\end{equation}

The operator of the total momentum of the system is
\begin{equation}
{\bf \hat{P}}_{c}={\bf \hat{p}}_{1}+{\bf \hat{p}}_{2}=-i\hbar {\bf %
\nabla }_{{\bf R}}\,.  \label{E51}
\end{equation}

By substituting operators~(\ref{E46})-(\ref{E49}) to the Hamilton function~(%
\ref{E18}), we get the nonrelativistic wave equation for a two-particle
system:
\begin{equation}
i\hbar \frac{\partial }{\partial t}\Psi ({\bf r,R,}t)=\left[ -\frac{\hbar
^{2}}{2M}\Delta _{{\bf R}}-\frac{\hbar ^{2}}{2\mu }\Delta _{{\bf r}}+V\bigl[%
r(1-\varepsilon )\bigr]\right] \Psi ({\bf r,R,}t)\,.  \label{E52}
\end{equation}
In this case, $\varepsilon $ is defined according to~(\ref{E45}), and $\mu $
is the reduced mass of the system, $\mu ={\displaystyle{\frac{m_{1\,}m_{2}}{%
m_{1}+m_{2}}}}$.

For the Hamiltonian $H$ not depending explicitly on time, the substitution
\begin{equation}
\Psi =\psi \exp \left( -i\frac{Et}{\hbar }\right) \,,  \label{E53}
\end{equation}
where $\psi $ depends on coordinates in the configuration space but not on
time, implies the nonlinear system of integro-differential equations for
stationary states of the two-particle system:
\begin{eqnarray}
&&\left[ -\frac{\hbar ^{2}}{2M}\Delta _{{\bf R}}-\frac{\hbar ^{2}}{2\mu }%
\Delta _{{\bf r}}+V\bigl[r(1-\varepsilon )\bigr] \right] \psi ({\bf r,R)=}%
E\psi ({\bf r,R)\,,}  \label{E54} \\
&&\varepsilon =\frac{\left\langle \psi ,\hat{\varepsilon}(F_{12}^{2})\psi
\right\rangle }{\left\langle \psi ,\psi \right\rangle }\,.  \label{E55}
\end{eqnarray}

By the substitution $\psi ({\bf r,R)=}\Phi {\bf (R)\,\varphi (r)}$, we can
separate motions of the center of masses of the system as a whole. As a
result, we obtained the following nonlinear system of equations:
\begin{eqnarray}
&&\left[ -\frac{\hbar ^{2}}{2\mu }\Delta _{{\bf r}}+ V\bigl[r(1-\varepsilon )%
\bigr]\right] \varphi ({\bf r)}=E\varphi ({\bf r)\,,}  \label{E56} \\
&&\varepsilon =\frac{\left\langle \varphi ,\hat{\varepsilon}%
(F_{12}^{2})\varphi \right\rangle }{\left\langle \varphi ,\varphi
\right\rangle }.  \label{E57}
\end{eqnarray}

As in the Schr\"{o}dinger nonrelativistic theory, a wave function $\varphi (%
{\bf r)}$ should be continuous together with its partial derivatives of the
first order in the whole space and, in addition, be a bounded single-valued
function of its arguments.

As in the Schr\"{o}dinger theory, for particles interacting by means of a
centrally symmetric potential, which depends only on the distance between
particles, the wave function $\varphi ({\bf r)}$ can be represented in the
following form:
\begin{equation}
\varphi ({\bf r)=}\frac{1}{r}\chi _{l}(r)Y_{lm}\left( \frac{{\bf r}}{r}%
\right) \,,  \label{E58}
\end{equation}
where $Y_{lm}\left( {\displaystyle{\frac{{\bf r}}{r}}}\right) $ are
orthonormalized spherical functions. Then the function $\chi _{l}(r)$
satisfies the following system of equations:
\begin{eqnarray}
&&\left[ -\frac{\hbar ^{2}}{2\mu }\left( \frac{d^{2}}{dr^{2}}-\frac{l(l+1)}{%
r^{2}}\right) +V\bigl[r(1-\varepsilon )\bigr]\right] \chi _{l}(r)=E\chi
_{l}(r)\,{\bf ,}  \label{E59} \\
&&\varepsilon =\frac{\left\langle \chi _{l},\hat{\varepsilon}%
(F_{12}^{2})\chi _{l}\right\rangle }{\left\langle \chi _{l},\chi
_{l}\right\rangle }\,.  \label{E60}
\end{eqnarray}

The quantity ${\displaystyle{\frac{m_{2}}{M}}}\varepsilon $ presents the
share of the momentum transferred to the second particle on measuring the
coordinate of the first. To construct the operator $\hat{\varepsilon}%
(F_{12}^{2})$ on the basis of classical mechanics is an extremely difficult
problem since one must be able to solve three-particle problems in the
general form. Therefore, by taking into account the properties of the
operator $\hat{\varepsilon}(F_{12}^{2})$ at $F_{12}\rightarrow 0$ $(\hat{%
\varepsilon}\rightarrow 0)$ and at $F_{12}\rightarrow \infty $ $(\hat{%
\varepsilon}\rightarrow 1)$, it is convenient to approximate $1-\hat{%
\varepsilon}$ in the first approximation by a gaussoid:
\begin{equation}
\hat{\varepsilon}=1-\exp \left( -\Omega _{0}F_{12}^{2}\left( \left| {\bf
\hat{r}}_{2}-{\bf \hat{r}}_{1}\right| \right) \right) \,,  \label{E61}
\end{equation}
where $\Omega _{0}$ is a parameter with dimensionality inversely
proportional to the square of force. The explanation for the definition of
this will be considered below.

The nonlinear system of the integro-differential equations~(\ref{E59}) - (%
\ref{E61}) for a two-particle system has solutions for definite values of
the energy $E$, but solutions with different energies are not orthogonal to
one another.

Moreover, since the parameter $\varepsilon $ is a quantum-mechanical mean in
every quantum state, we may say that every quantum state has its own
potential of interaction between particles.

The constant $\Omega _{0}$ can be defined by analyzing the discrete spectrum
of a hydrogenlike atom. Let two particles with masses $m_{1}$ (electron) and
$m_{2}$ (atomic nucleus) be coupled through the Coulomb potential $V(r)=-{%
\displaystyle{\frac{Ze^{2}}{r}}}$, where $Z$ is the charge of the atomic
nucleus. The nonlinear system of integro-differential equations for bound
states~(\ref{E59}) - (\ref{E61}) can be written in the following way:
\begin{eqnarray}
&&\,\left[ -\frac{\hbar ^{2}}{2\mu }\left( \frac{d^{2}}{dr^{2}}-\frac{l(l+1)%
}{r^{2}}\right) -\frac{Ze^{2}}{r(1-\varepsilon _{{\rm nl}})}\right] \chi _{%
{\rm nl}}(r)=E_{{\rm nl}}\chi _{{\rm nl}}(r)\,{\bf ,}  \label{E62} \\
&&\varepsilon _{{\rm nl}}=1-\int_{0}^{\infty }\chi _{{\rm nl}}^{2}(r)\exp
\left( -\Omega _{0}\frac{Z^{2}e^{4}}{r^{4}(1-\varepsilon _{{\rm nl}})^{4}}%
\right) \,dr\,,  \label{E63} \\
&&\int_{0}^{\infty }\chi _{{\rm nl}}^{2}(r)\,dr=1\,.  \label{E64}
\end{eqnarray}
Here, $\mu $ is the reduced mass of the system.

Equations~(\ref{E62}) and (\ref{E64}) define normalized radial functions of
a hydrogenlike atom by the Schr\"{o}dinger theory. Their solutions for bound
states are well known (see, e.g.,~\cite{R6}):
\begin{equation}
\chi _{{\rm nl}}(r)=N_{{\rm nl}}r^{l+1}F\left( -n+l+1,2l+2,\frac{2Zr}{%
(1-\varepsilon _{{\rm nl}})na_{0}}\right) \exp \left( \frac{Zr}{%
(1-\varepsilon _{{\rm nl}})na_{0}}\right) \,,  \label{E65}
\end{equation}
where
\begin{equation}
N_{{\rm nl}}=\frac{1}{(2l+1)!}\left[ \frac{(n+l)!}{2n(n-l-1)!}\right]
^{1/2}\left( \frac{2Z}{(1-\varepsilon _{{\rm nl}})na_{0}}\right) ^{l+3/2}\,.
\label{E66}
\end{equation}
Here, $a_{_{0}}={\displaystyle{\frac{\hbar ^{2}}{\mu e^{2}}}}$ is the Bohr
radius, and $F$ is a degenerate hypergeometric function.

Eigenvalues of energy are expressed as
\begin{equation}
E_{{\rm nl}}=-\frac{\mu c^{2}}{2}\frac{(\alpha Z)^{2}}{n^{2}}\frac{1}{%
(1-\varepsilon _{{\rm nl}})^{2}}\,.  \label{E67}
\end{equation}
Here, $\alpha ={\displaystyle{\frac{e^{2}}{\hbar c}}}$ is the fine structure
constant, $l=0,1,\ldots ,n-1,$ and $n=1,2,\ldots ,\infty $.

By substituting $\chi _{{\rm nl}}(r)$ into Eq.~(\ref{E63}), we obtain the
nonlinear equation for the determination of $\varepsilon _{{\rm nl}}$:
\begin{equation}
\eta _{{\rm nl}}=S_{{\rm nl}}\int_{0}^{\infty }x^{{\rm 2l+2}}\exp \left( -x-%
\frac{\Omega (\alpha Z)^{6}}{\eta _{{\rm nl}}^{8}n^{4}x^{4}}\right)
F^{2}\left( -n+l+1,2l+2,x\right) \,dx\,,  \label{E68}
\end{equation}
where $\eta _{{\rm nl}}=1-\varepsilon _{{\rm nl}}$ , $S_{{\rm nl}}={%
\displaystyle{\frac{1}{\left[ (2l+1)!\right] ^{2}}}}\left[ {\displaystyle{%
\frac{(n+l)!}{2n(n-l-1)!}}}\right] $ and $\Omega ={{{\displaystyle{\frac{%
16\Omega _{0}(\mu c^{2})^{4}}{\hbar ^{2}c^{2}}}}}}$.

For the ground state of a hydrogenlike atom, we have
\begin{equation}
\eta _{10}=\frac{1}{2}\int_{0}^{\infty }x^{2}\exp \left( -x-\frac{\Omega
(\alpha Z)^{6}}{\eta _{10}^{8}\,x^{4}}\right) \,dx\,.  \label{E69}
\end{equation}
For $\eta _{10}$, the nonlinear Eq.~(\ref{E69}) has solutions if $\Omega
(\alpha Z)^{6}\leq \Omega _{\text{{\it c}}}=0.40765$ that is depicted in
Fig.~\ref{fig1}. From two solutions, that solution is considered as suitable
which is located nearer to 1. The second solution should be omitted. Indeed,
it corresponds to the case where $\varepsilon $ tends to 1 as the parameter $%
\alpha Z$ decreases to zero, which contradicts to the assumptions made above
about the parameter of noncommutativity. For $\Omega (\alpha Z)^{6}>\Omega _{%
\text{{\it c}}}$, Eq.~(\ref{E69}) has no solutions, which means it is
impossible for a given bound state to exist.

The system of Eqs.~(\ref{E62})-(\ref{E64}) describes a nonrelativistic
motion of a particle with mass $\mu $ in the external field $V(r)=-{%
\displaystyle{\frac{Ze^{2}}{r}}}$ and possesses a critical value of the
constant $Z=Z_{\text{{\it c}}}$ such that a given bound state cannot exist
if it is exceeded. The relativistic equation for a hydrogenlike atom is the
Dirac equation~\cite{R7} for a particle with mass $\mu $ in the Coulomb
field, and it also has a critical constant of the ground state equal to $Z_{%
\text{{\it c}}}=1/\alpha $. It is reasonable to assume that these constants
are the same. This presents the possibility to determine the constant $%
\Omega _{0}$ as
\begin{equation}
\Omega _{0}=\Omega _{\text{{\it c}}}\frac{\hbar ^{2}c^{2}}{16(\mu c^{2})^{4}}%
.  \label{E70}
\end{equation}

In Fig.~\ref{fig2}, we plot values of the binding energy for the ground
state of a hydrogenlike atom, which is calculated by using the parameter $%
\Omega _{0}$ defined in such a way. There, we also present the analogous
binding energies according to Schr\"{o}dinger and Dirac. It is seen that the
results of our calculation occupy the intermediate position. Similar
calculations can be easily performed for excited levels of hydrogenlike
atoms. In this case, levels by Schr\"{o}dinger with a given $n$ split into $%
n $ close sublevels since the orbital quantum number $l$ can take $n$ values
$(l=0,1,\ldots ,n-1)$, i.e., the degeneration is removed in this case. All
levels with given $n$ and different $l$ are located under the corresponding
Schr\"{o}dinger level. The value of the nonrelativistic splitting is much
less than that calculated by the Dirac theory. The parameter of
noncommutativity for the operators of coordinates and momenta of different
particles $\varepsilon $, presented in Fig.~\ref{fig3}, increases with the
quantum numbers $n$ and $l$ (for the same $Z$). That is, fully
nonrelativistic solutions transfer to that of the Schr\"{o}dinger equation
for large quantum numbers.

As a peculiar feature of a fully nonrelativistic equation, we indicate the
presence of a critical value of the parameter $\alpha Z_{\text{{\it c}}}$
for any energy level that is not revealed by the Schr\"{o}dinger
nonrelativistic equation. For example, if $n=2$, $\alpha Z_{\text{{\it c}}}=3
$ for $l=0$ and $\alpha Z_{\text{{\it c}}}=2.5$ for $l=1$.

When the parameter $\alpha Z$ grows, the mean distance between particles
decreases. For the ground state of a hydrogenlike atom, it is defined as
\begin{equation}
\langle |{\bf \hat{r}}_{2}-{\bf \hat{r}}_{1}|\rangle =\frac{3\hbar }{2\mu c}%
\cdot \frac{(1-\varepsilon _{10})^{2}}{\alpha Z}  \label{E71}
\end{equation}
and takes the smallest value equal to $\langle |{\bf \hat{r}}_{2}-{\bf \hat{r%
}}_{1}|\rangle \approx 0.9{\displaystyle{\frac{\hbar }{\mu c}}}$ when $%
\alpha Z=1$. At the same time (i.e., for $\alpha Z=1$), mean distances
between particles significantly exceed the value ${\displaystyle{\frac{\hbar
}{\mu c}}}$ for excited quantum states ($\langle |{\bf \hat{r}}_{2}-{\bf
\hat{r}}_{1}|\rangle \approx 5.9{\displaystyle{\frac{\hbar }{\mu c}}}$ for $%
n=2$ and $l=0$; $\langle |{\bf \hat{r}}_{2}-{\bf \hat{r}}_{1}|\rangle
\approx 5{\displaystyle{\frac{\hbar }{\mu c}}}$ for $n=2$ and $l=1$). If the
parameter $\alpha Z$ further grows, the nonlinear system of Eqs.~(\ref{E62}%
)-(\ref{E64}) has no solutions for the state with $n=1$, i.e., the $1{\rm S}$
state cannot exist, and the ground state is a state with $n=2$ and $l=0$ ($2%
{\rm S}$ state) or $l=1$ ($2{\rm P}$ state). It is possible if
$1<\alpha Z<3$. Here we are faced with the essential difference
from solutions of the Schr\"{o}dinger nonrelativistic equation,
for which, as is well known, the ground state is always the $1{\rm
S}$ state.

Below, we present the quantum Poisson brackets introduced by Dirac~\cite{R7}
as
\begin{eqnarray}
&&\{\hat{x}_{1},\,\hat{p}_{1x}\}=1-\frac{m_{2}}{M}\varepsilon \,,
\label{E72} \\
&&\{\hat{x}_{2},\,\hat{p}_{2x}\}=1-\frac{m_{1}}{M}\varepsilon \,,
\label{E73} \\
&&\{\hat{x}_{1},\,\hat{p}_{2x}\}=\frac{m_{2}}{M}\varepsilon \,,  \label{E74}
\\
&&\{\hat{x}_{2},\,\hat{p}_{1x}\}=\frac{m_{1}}{M}\varepsilon \,,  \label{E75}
\\
&&\{\hat{x}_{1},\,\hat{x}_{2}\}=0\,,  \label{E76} \\
&&\{\hat{p}_{1x},\,\hat{p}_{2x}\}=0\,.  \label{E77}
\end{eqnarray}

For $\varepsilon \rightarrow 0$, these brackets transfer to the classical
Poisson brackets, i.e., we have the complete analogy between classical and
quantum mechanics in this case. As is seen in Fig.~\ref{fig3}, $\varepsilon $
is remarkably different from zero for systems whose sizes are of the order
of the Compton wavelength of particles forming the system. In this case,
there is no analogy with classical mechanics. To what extent it would take
place can be judged by comparing the proposed theory with experiment. But it
is already clear that we have obtained a considerably better agreement with
the result of solving the relativistic Dirac equation for the ground state
of a hydrogenlike atom as compared with the Schr\"{o}dinger nonrelativistic
theory.

\section{A nonrelativistic system of $N$-interacting particles}

The previous results can be easily generalized to a system consisting of $N$
particles interacting among themselves via two-particle forces. Let the
operators of the coordinates and momenta of $N$ particles be ${\bf \hat{r}}%
_{1},{\bf \hat{r}}_{2},\,\ldots \,,{\bf \hat{r}}_{N}$, ${\bf \hat{p}}_{1},%
{\bf \hat{p}}_{2},\,\ldots \,,{\bf \hat{p}}_{N}$. Define the operators of
coordinates and momentum of the center of masses of the system:
\begin{eqnarray}
&&{\bf \hat{r}}_{\text{{\it c}}}=\frac{1}{M}\sum_{k=1}^{N}m_{k}{\bf \hat{r}}%
_{k}\,,  \label{E78} \\
&&{\bf \hat{P}}_{\text{{\it c}}}=\sum_{k=1}^{N}{\bf \hat{p}}_{k}\,.
\label{E79}
\end{eqnarray}
Here, $M=\sum_{k=1}^{N}m_{k}$ is the mass of the whole system.

By analogy with a two-particle problem, we require that the commutator of
the operator of the coordinate for any particle with the operator of the
total momentum of the system be equal to $i\hbar $:

\begin{equation}
\left[ \hat{x}_{k},\,\hat{P}_{\text{{\it c}}x}\right] =\left[ \hat{y}_{k},\,%
\hat{P}_{\text{{\it c}}y}\right] =\left[ \hat{z}_{k},\,\hat{P}_{\text{{\it c}%
}z}\right] =i\hbar \text{\quad }(k=1,\,2\,,\,\ldots \,,N).  \label{E80}
\end{equation}
Then, if

\begin{equation}
\left[ \hat{x}_{j},\hat{p}_{kx}\right] =\left[ \hat{y}_{j},\hat{p}_{ky}%
\right] =\left[ \hat{z}_{j},\hat{p}_{kz}\right] =i\hbar \frac{m_{k}}{M}%
\varepsilon _{jk}\quad (\ j\neq k),  \label{E81}
\end{equation}
we have
\begin{equation}
\left[ \hat{x}_{j},\hat{p}_{jx}\right] =\left[ \hat{y}_{j},\hat{p}_{jy}%
\right] =\left[ \hat{z}_{j},\hat{p}_{jz}\right] =i\hbar \left[
1-\sum_{k=1}^{N}\frac{m_{k}}{M}\varepsilon _{jk}\right] \,,\text{\quad }%
j=1,\,2,\,\ldots \text{\thinspace },N.  \label{E82}
\end{equation}
Here, the parameter of noncommutativity of the operators of coordinates and
momenta of different particles
\begin{equation}
\varepsilon _{jk}=\frac{\left\langle \psi ,\hat{\varepsilon}(F_{jk}^{2},\mu
_{jk})\psi \right\rangle }{\left\langle \psi ,\psi \right\rangle }\,
\label{E83}
\end{equation}
is symmetric with respect to a permutation of the indices $j,k$ and
identically equals zero for $j=k$ by definition.

In addition,
\begin{equation}
\left[ \hat{x}_{j},\hat{x}_{k}\right] =\left[ \hat{y}_{j},\hat{y}_{k}\right]
=\left[ \hat{z}_{j},\hat{z}_{k}\right] =0\,,\text{\quad }\left[ \hat{p}_{jx},%
\hat{p}_{kx}\right] =\left[ \hat{p}_{jy},\hat{p}_{ky}\right] =\left[ \hat{p}%
_{jz},\hat{p}_{kz}\right] =0\,.  \label{E84}
\end{equation}

One of the possible representations for the operators of coordinates and
momenta of particles can be written as
\begin{eqnarray}
&&{\bf \hat{r}}_{j}={\bf r}_{j}\,,\text{\quad }j=1,\,2,\,\ldots \,,N\,\,,
\label{E85} \\
&&{\bf \hat{p}}_{j}=-i\hbar \left[ 1-\sum_{q=1}^{N}\frac{m_{q}}{M}%
\varepsilon _{jq}\right] {\bf \nabla }_{j}-i\hbar \sum_{k=1}^{N}\frac{m_{j}}{%
M}\varepsilon _{jk}{\bf \nabla }_{k}\,,\text{\quad }j=1,\,2,\,\ldots \,,N\,.
\label{E86}
\end{eqnarray}
Here, we take coordinates of particles as independent variables since the
corresponding operators mutually commute.

In this case, a nonlinear system of equations for the nonrelativistic
problem of $N$ particles has the form
\begin{equation}
\left\{ -\frac{\hbar ^{2}}{2}\sum_{i=1}^{N}\left[ \frac{A_{i}}{m_{i}}\Delta
_{i}+\sum_{k>i}^{N}\frac{2B_{ik}}{M}({\bf \nabla }_{i}\cdot {\bf \nabla }%
_{k})\right] +\sum_{i=1}^{N}\sum_{j>i}^{N}V(\left| {\bf r}_{j}-{\bf r}%
_{i}\right| )\right\} \Psi =E\Psi \,,  \label{E87}
\end{equation}
where
\begin{eqnarray}
&&A_{i}=\left( 1-\sum_{q=1}^{N}\frac{m_{q}}{M}\varepsilon _{iq}\right)
^{2}+\sum_{q=1}^{N}\frac{m_{i\,}m_{q}}{M^{2}}\varepsilon _{iq}^{2}\,,
\label{E88} \\
&&B_{ik}=\left( 2-\sum_{q=1}^{N}\frac{m_{q}}{M}(\varepsilon
_{iq}+\varepsilon _{kq})\right) \varepsilon _{ik}+\sum_{q=1}^{N}\frac{m_{q}}{%
M}\varepsilon _{iq}\varepsilon _{kq}\,.  \label{E89}
\end{eqnarray}
Here, $\varepsilon _{ik}$ is defined according to~(\ref{E83}), and
\begin{equation}
\hat{\varepsilon}(F_{jk}^{2},\mu _{jk})=1-\exp \left( -\Omega _{{\it c}}{{%
\frac{\hbar ^{2}c^{2}}{16(\mu _{jk}c^{2})^{4}}}}F_{jk}^{2}(\left| {\bf \hat{r%
}}_{j}-{\bf \hat{r}}_{k}\right| )\right) \,,  \label{EE}
\end{equation}
$\Omega _{c}=0.40765$, and $\mu _{jk}={\displaystyle{\frac{m_{j}m_{k}}{%
m_{j}+m_{k}}}}$.

It can be shown that the introduction of the so-called Jacobi coordinates
provides the separation of motion of the center of masses as a whole.

The system of equations~(\ref{E87})-(\ref{E89}) takes a particularly simple
form in the important case of identical particles $(m_{j}=m,\quad
\varepsilon _{jk}=\varepsilon ,$ $\ \ j,k=1,2,\,\ldots \,,N\;,$ $\ j\neq k)$
after the introduction of the so-called normed Jacobi coordinates,
\begin{eqnarray}
&&{\bf q}_{k}=\sqrt{\frac{k}{k+1}}\left( \frac{1}{k}\sum_{s=1}^{k}{\bf r}%
_{s}-{\bf r}_{k+1}\right) \,,\text{ \ \ \ \ }1\leq k\leq N-1,  \label{E90} \\
&&{\bf q}_{N}=\sqrt{\frac{1}{N}}\sum_{s=1}^{N}{\bf r}_{s}\,.  \label{E91}
\end{eqnarray}

In this case, after the separation of motion of the center of masses, we
have
\begin{eqnarray}
&&\left\{ -\frac{\hbar ^{2}(1-\varepsilon )^{2}}{2m}\left( \Delta _{{\bf q}%
_{1}}+\cdots +\Delta _{{\bf q}_{N-1}}\right) +V({\bf q}_{1},\;\ldots \;,{\bf %
q}_{N-1})\right\} \varphi =E\varphi \,,\,  \label{E92} \\
&&\varepsilon =\frac{\left\langle \varphi ,\hat{\varepsilon}(F_{12}^{2},%
\frac{m}{2})\varphi \right\rangle }{\left\langle \varphi ,\varphi
\right\rangle }\,.  \label{E93}
\end{eqnarray}
Here, $V({\bf q}_{1},{\bf q}_{2},\,\ldots \,,{\bf q}_{N-1})$ represents the
potential energy of the interaction of all particles expressed in terms of
Jacobi coordinates~(\ref{E90}).

Since $0\leq \varepsilon <1$, the mean value of the kinetic energy will be
less than that according to Schr\"{o}dinger, and the energies of the bound
states will be situated lower than the Schr\"{o}dinger ones.

\section{Conclusion}

The Schr\"{o}dinger equation for a system of interacting particles is not a
strictly nonrelativistic equation because it is grounded on the implicit
assumption about finiteness of the interaction propagation velocity. The
last means that if the commutator of operators of a coordinate and the
corresponding momentum of a free particle is defined as
\begin{equation}
\left[ \hat{x},\,\hat{p}_{x}\right] =i\hbar \,,  \label{E94}
\end{equation}
this commutator for a system of coupled particles has the same value $i\hbar
\,$.$\,$However, in a nonrelativistic quantum system during measurement of
the coordinate of a particle, a whole transferred momentum is distributed
over all particles but is not transferred to only the measured one.
Therefore, in a system of interacting particles, this commutator should have
the form
\begin{equation}
\left[ \hat{x},\,\hat{p}_{x}\right] =i\hbar \delta \,,  \label{E95}
\end{equation}
where $0 < \delta \leq 1$ .

The rejection of the implicit assumption on finiteness of the propagation
velocity of interactions implies the noncommutativity of the operators of
coordinates and momenta of different particles. But the operators of
coordinates of all particles and operators of momenta of all particles
mutually commute that allows one to use these collections as independent
variables.

The deduced nonlinear system of integro-differential equations allows one to
separate the motion of the center of masses of the system which moves as a
free particle.

A solution of this essentially nonlinear system exists for completely
definite values of the energy of the system. The wave functions
corresponding to these energies, as a rule, are mutually nonorthogonal.

Properties of solutions of the proposed nonlinear system of equations
essentially differ from Schr\"{o}dinger solutions for systems, for which the
Compton wavelength of particles is comparable with the size of the system.
That is, the consideration of noncommutativity of the operators of
coordinates and momenta of different particles is of importance for quantum
mechanics of atoms with large charge of the nucleus $(\alpha Z\sim 1)$ and
for phenomena of the physics of atomic nuclei, where the size of the system
is of the order of the Compton wavelength of particles which compose the
system.

The author acknowledges Dr.~V.V.~Kukhtin and Dr.~A.I.~Steshenko for a very
fruitful discussion.

\begin{figure}[tbp]
\caption{Dependence of the right side of Eq. (69) on $\protect\eta $ for
various values of the parameter $\Omega (\protect\alpha Z)^{6}.$}
\label{fig1}
\end{figure}

\begin{figure}[tbp]
\caption{Binding energy of the ground state of a hydrogenlike atom [Eq.(67)]
vs the parameter $\protect\alpha Z$. The upper dotted line corresponds to
the Schr\"{o}dinger theory, $E_{S}=-{\displaystyle{\frac{\protect\mu c^{2}(%
\protect\alpha Z)^{2}}{2}}}$, and the lower one to the Dirac theory, $E_{D}=%
\protect\mu c^{2}\left\{ -1+[1-(\protect\alpha Z)^{2}]^{1/2}\right\} $.}
\label{fig2}
\end{figure}

\begin{figure}[tbp]
\caption{Dependence of the parameter of noncommutativity of operators $%
\protect\varepsilon $ on the parameter $\protect\alpha Z$ for the lowest
states of a hydrogenlike atom.}
\label{fig3}
\end{figure}

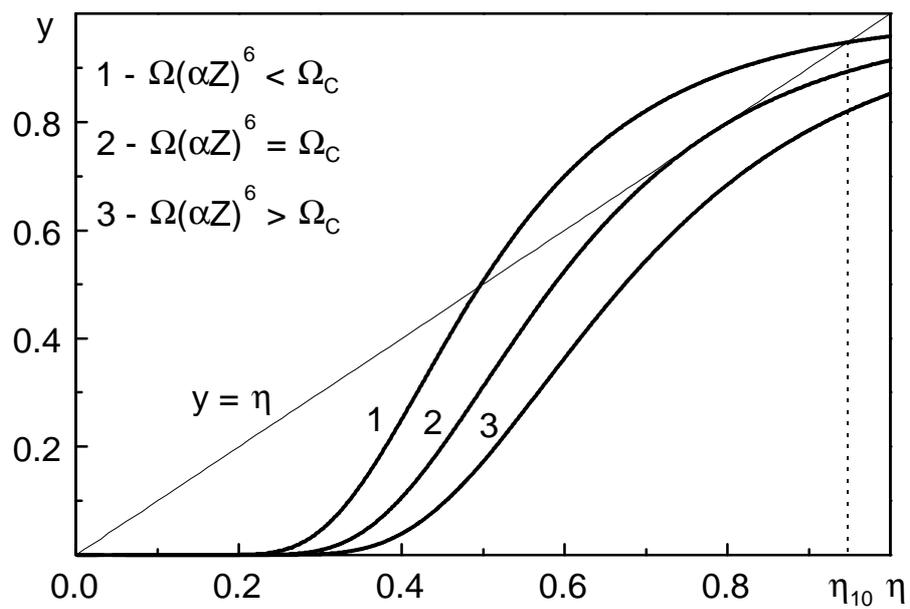
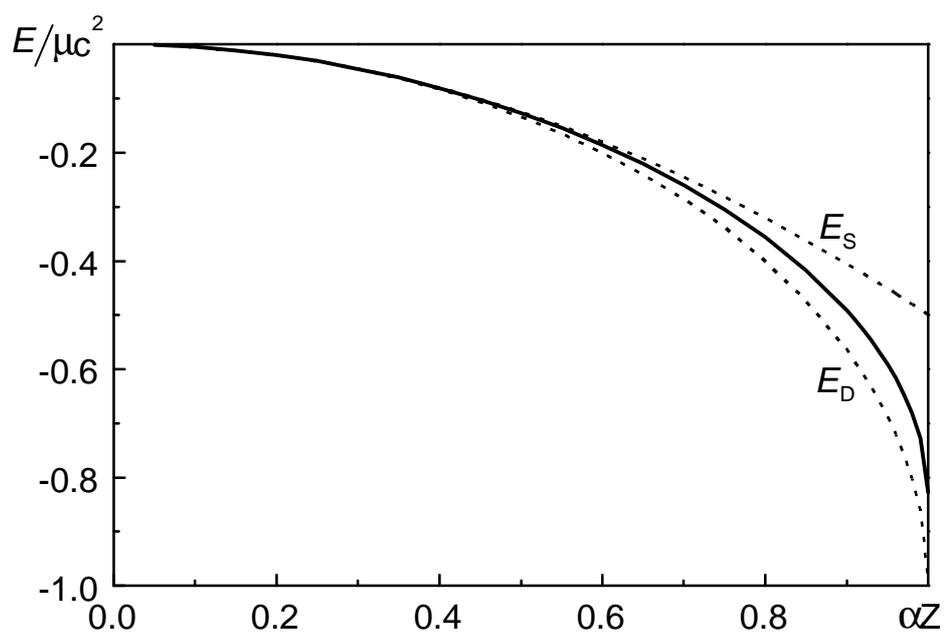
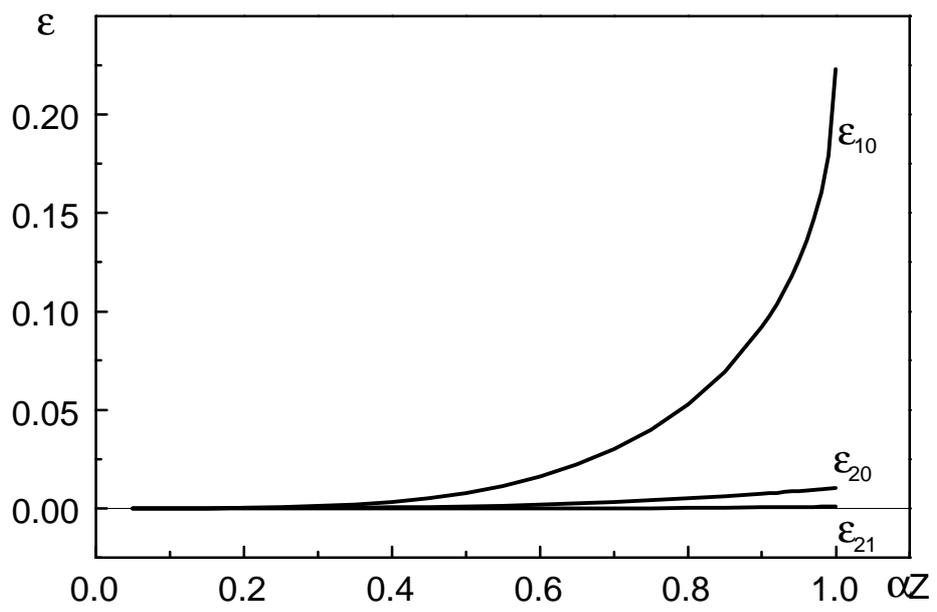
\end{document}